\documentclass[conference]{IEEEtran}
\IEEEoverridecommandlockouts
\usepackage{times}
\usepackage{epsfig}
\usepackage{graphicx}
\usepackage{amsmath}
\usepackage{amssymb}
\usepackage{multirow}
\usepackage{booktabs}
\usepackage{url}
\usepackage[table]{xcolor}
\usepackage{xcolor}
\usepackage{pgfplotstable}
\usepackage[left=0.75in, right=0.75in, top=1in, bottom=0.75in]{geometry}
\usepackage{url}
\usepackage{float}

\begin{document}
	

	\title{ 
		Intelligent Stress Assessment for e-Coaching 
		\thanks{}
	}
	
	\author{\IEEEauthorblockN{Kenneth Lai\textsuperscript{1,2}, Svetlana Yanushkevich\textsuperscript{1}, Vlad Shmerko\textsuperscript{1}}
		\IEEEauthorblockA{\textsuperscript{1}Biometric Technologies Laboratory, Department of Electrical and Software Engineering, University of Calgary, Canada\\ 
			\IEEEauthorblockA{\textsuperscript{2}Department of Clinical Neurosciences, Cumming School of Medicine, University of Calgary, Canada\\ 
				Email: \{kelai, syanshk, vshmerko\}@ucalgary.ca}}
	}


	
\markboth{IEEE SSCI,~2023}{ \MakeLowercase{\textit{et al.}}:
	.....}	

	\maketitle
	
\begin{abstract}
	This paper considers the adaptation of the e-coaching concept at times of  emergencies and disasters, through aiding the e-coaching with intelligent tools for  monitoring humans' affective state. The states such as anxiety, panic, avoidance, and stress, if properly detected, can be mitigated using the e-coaching tactic and strategy. In this work, we focus on a stress monitoring assistant tool developed on machine learning techniques. We provide the results of an experimental study using the proposed method. 
\end{abstract}

\begin{IEEEkeywords}
	Affective Computing, E-Coaching, Stress Assessment, Deep Learning, Machine Reasoning.
\end{IEEEkeywords}

\section{Introduction}\label{sec:}

The adaptation of e-coaching in terms of the community's mental health is an ongoing process \cite{[Varma-2021],[Yan-2021]}. As stated in \cite{[Andre-2021]}, e-coaching ``may contribute to a better understanding of people's affective responses to the COVID-19 crisis. If ethical, legal, and social implications are addressed appropriately, affective computing technologies may bring a real benefit to society by monitoring and improving people's mental health''.  Typical symptoms include anxiety, panic, avoidance, and stress. In particular, monitoring of COVID-19-related stress globally in 63 countries has shown that over 70\% of the respondents had greater than moderate levels of stress, with 59\% meeting the criteria for clinically significant anxiety and 39\% reporting moderate depressive symptoms \cite{[Varma-2021]}. At the time of writing this paper, climate change and armed conflicts cause mass migration and related disastrous consequences to human mental health.


In our study, we suggest that e-coaching should be integrated into a system developed to manage global disaster events \cite{[WHO-2020],[Joshi-2019]}. This system is known as Emergency Management Cycle \cite{[WHO-EMC]}. This  is adapted in our study since it provides a systematic counter-disaster view of e-coaching. 
The framework proposed in this paper paves the way to a strategic road mapping for  e-coaching technologies in the era of natural and human-made disasters. This framework is based on the three technology-society premises.

\emph{First premise:}
E-coaching resources must be integrated into the standardized four-phase mechanism \cite{[WHO-EMC]}: prevent, prepare, respond, and recover phase. 

\emph{Second premise:}
When designing the e-coaching component, a risk mitigation mechanism should be integrated into the stress-conditional scenarios which are typical in pandemics \cite{[Varma-2021]}. In other words, the stress detector should continuously learn the user's stress pattern in order to adjust the e-coaching tactic and strategy.

\emph{Third premise:} 
Typically, an e-coaching system is viewed as a network of wearable and wireless sensors. Some of them can be utilized for stress detection which has traditionally been a part of the affect recognition process \cite{[Schmidt-2019]} which includes the detection of emotional states such as sadness, happiness, and surprise. Hence, an experiment must be set up in order to determine what kind of sensors are useful for this purpose. The measure of usefulness includes accuracy, among others. 

\section{Two-stage e-Coaching Technology}

\begin{figure}[!ht]
	\begin{center}
		\includegraphics[width=0.3\textwidth]{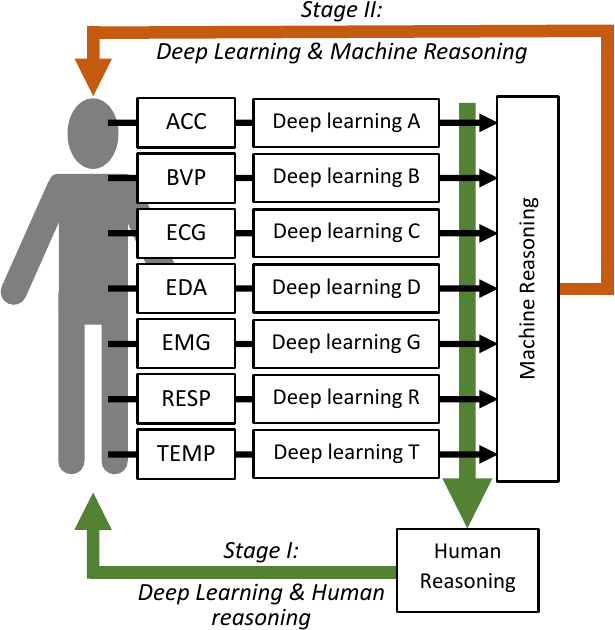}
	\end{center}
	\caption{Illustration of a two-stage application of intelligent tools.}
	\label{fig:Echelon-processing}
\end{figure}

In this paper, two-stage intelligent processing, as seen in Fig. \ref{fig:Echelon-processing}, is used:
\begin{itemize}
	\item \emph{Stage I} is aimed at gathering physiological information from a subject for human decision-making (reasoning).
	\item \emph{Stage II} is aimed at supporting the human decision-maker via machine reasoning.
\end{itemize}

Note that the two-stage intelligent processing has been efficiently used. For example, in fish classification \cite{[Abinaya-2021]}, after the automatic segmentation of fish images into head, scales, and body, features are extracted from each segment using a deep learning network. A Bayesian network served as a feature fusion tool. An efficient combination of these computational intelligence techniques was reported in \cite{[Lera-2017]} for context awareness in autonomous robots (deep profiling environmental sounds aimed at the improvement of context recognition). In \cite{[Yang-2019]}, video sequences are processed using deep learning of frames with Bayesian inference of depth estimates between different time frames. Paper \cite{[Chen-2018]} reports the results of regular inspection videos for identifying cracks in nuclear power plant components using a deep learning approach and Bayesian network.

\section{Stage I: Deep profiling}
\label{sec:Proof-of-concept}

In this section, we describe the experiments for the first stage of intelligent processing explained in Fig. \ref{fig:Echelon-processing}. In our experiment, we adopted the concept of continuous stress monitoring for first responders \cite{[Lai-2021]} for e-coaching users. 

The public-centric and occupational stresses have traditionally been differentiated in terms of their detection, monitoring, and responses.
 Occupational stresses have been described in the standards for decision support \cite{dhsCOVID19}. A Wearable Sensor Network is a \textbf{preferable} tool for stress detection and monitoring \cite{dias2018wearable} in the workplace and occupational hazard monitoring scenarios \cite{zhang2020video}. 

\begin{figure}[!ht]
	\begin{center}
		\includegraphics[width=0.49\textwidth]{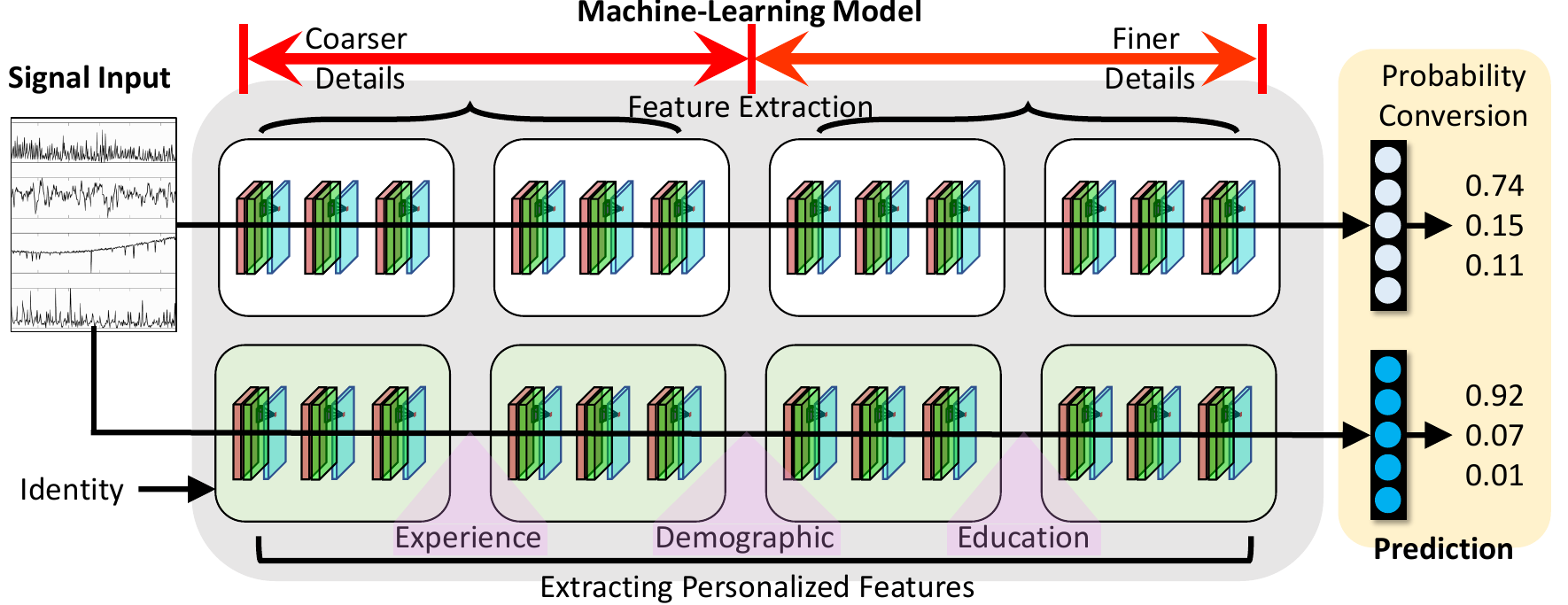}
	\end{center}
	\caption{The core of stress monitoring assistants is the deep learning network of the Res-TCN architecture. The top portion of the network has a general design, while the bottom network is built for personalized processing which provides the subject's identity information which improves the model's performance.}
	\label{}
\end{figure}

\subsection{Experimental scenario}
In our experiment, we assume the availability of physiological signals supplied by the Wearable Sensor Network, e.g., ECC, EDA, BVP, etc. In our study, we divide the stress states into Yes (high level) or No (low level), as well as  the stress state is recognized among other emotional states. The primary goal of our experiments is to demonstrate that continuous stress monitoring in e-coaching has been advanced from the category of a 'working idea' to the category of 'prototyping'.

\subsection{Dataset}
 In modeling, in general, synthetic data should replicate the real data as close as possible, the sample sizes must satisfy the criteria of statistical significance, and a standard protocol must be followed in order to guarantee the repeatability of the experiment. A dataset that partially satisfies the above requirements is the WESAD dataset, Multimodal Dataset for Wearable Stress and Affect Detection by \emph{Schmidt et al.} \cite{schmidt2018introducing}. 

The WESAD dataset was collected from 17 participants, each wearing seven sensors (ACC, ECG, BVP, EDA, EMG, RESP, and TEMP). For each signal different partition is labeled by one of the four different affective states: neutral, stressed, amused, and meditated. There are four different test scenarios: normal, amusement, stress, and meditation. The neutral scenario lasted the first 20 minutes: the participants were asked to do normal activities such as reading a magazine and sitting/standing at a table. In the amusement scenario, the participants watched 11 funny video clips for a total length of 392 seconds. The stress scenario required the participants to perform public speaking and arithmetic tasks for a total of 10 minutes. The last scenario involved a guided meditation session of 7 minutes in duration. The ground truth labels for the affect states were collected using the Positive and Negative Affect Schedule (PANAS) scheme \cite{watson1988development}, upon completion of each trial.

\subsection{Measures}
To assess the classification algorithm performance, it is important to determine the most suitable performance indicators. In the case of balanced data, the traditional measures include:

\begin{itemize}
	\item []\hspace{-7.5mm} $TP$ -- True Positives (correct predictions of emotion),
	\item []\hspace{-7.5mm} $FN$ -- False Negatives (incorrect predictions of emotion),
	\item []\hspace{-7.5mm} $TN$ -- True Negatives (correct rejections of emotion), and 
	\item []\hspace{-7.5mm} $FP$ -- False Positives (incorrect predictions of emotion). 
\end{itemize}

These measures form a $2\times 2$ confusion matrix and are used to derive accuracy, recall, precision, receiver operating characteristics, and balanced $F_1$-score \cite{luque2019impact}. We selected a few for evaluation, i.e., the accuracy measure and $F_1$-score:
\begin{eqnarray*}
	\texttt{Accuracy}&=&\frac{\text{\textit{TP+TN}}}{\text{\textit{TP+FN+TN+FP}}}\label{eq:acc}\\
	\texttt{\small $F_1$-score}&=&2\times\frac{ \texttt{\small Precision} \times \texttt{\small Recall}}{\texttt{\small Precision} + \texttt{\small Recall}}\label{eq:f1}
\end{eqnarray*}
where \texttt{Recall} (also known as sensitivity) represents the system's ability to detect a specific emotion, $\texttt{Recall}={\textit{TP}}/{(\textit{TP+FN})}$, and \texttt{Precision} (also called positive predictive value) is the system's ability to be correct on a predicted emotion: $\texttt{Precision}={\textit{TP}}/{(\textit{FP+TP})}$.

Accuracy reflects the number of correctly classified patterns among the samples, and, thus, it is a \emph{probability of success} in recognizing the right class of an instance. However, in the case of highly imbalanced datasets, the accuracy measure (\ref{eq:acc}) is \textbf{misleading}. A classifier that is very effective in predicting the majority class but misses most of the minority classes may easily have very high accuracy \cite{maratea2014adjusted}.

The $F_1$-score is a weighted average (harmonic mean) of precision and recall rates, representing the system's balanced ability to detect a specific emotion correctly. The $F_1$-score reaches 1 at perfect precision and recall, and 0 at the worst of both \cite{luque2019impact}. This measure provides a way of combining the recall and the precision in order to capture both.

\subsection{Choosing a deep learning network}

For analyzing emotional states, we chose the Temporal Convolutional Network (TCN), Recurrent Neural Network (RNN). The TCN offers a solution to quickly learn patterns from time-series data; it consists of a series of causal 1D convolution layers optimized for sequential data. It was used for the classification of stress \cite{feng2019dynamic} and early predictions \cite{catling2020temporal}. 
In our work, the TCN is chosen for the task of emotion classification because of the following reasons:

1) \emph{Classification task}: Since the goal of this paper is to perform emotion classification and not image generation, the GANs are not suitable.

2) \emph{Time-series data}: Our input data are physiological signals such as ECG which are time-based data points, the TCN and RNN are best for such data types.

3) \emph{Time complexity}: Due to the nature of convolution in TCN, the process is consistent and easier to parallelize, as opposed to RNN which requires the previous step to be finished before performing the next operation.

4) \emph{Memory}: The TCN requires much less memory (parameters) compared to RNN when processing the long input sequences. In addition, the TCN can obtain a specific receptive field based on the number of residual blocks, while the RNN always uses the maximum length of the sequence.

\subsection{Experiment I: Identification of a stressed user in the e-coached team }

In this paper, we deploy the analysis of physiological signals in order to detect and identify the level of stress. 

Before analyzing the personalized data, we must determine whether the provided data can be used for subject identification, that is, given a sample of accelerometer data, can we identify the subject? This is a vital task for personalized stress detection as it links subjects to their corresponding stress levels. In Table \ref{tab:id}, we report the accuracy and $F_1$-score of using various physiological signals for subject identification. In this table, the performance is reported for 10 modalities, including 6 signals from the chest region and 4 from the wrist region. The performance measures are calculated using 10-fold cross-validation.

\subsubsection*{Observation 1 (Highest performance)}
The highest performance is obtained via the RESP signal collected from the chest region with an accuracy of 99.84\%. The next highest-performing signal is the BVP signal collected from the wrist region with an accuracy of 99.60\%.

These results suggest that it is possible to recognize the identity of the sensor wearer given these two types of signals. The least useful descriptors are the TEMP signal from both the chest and wrist sensors. Analysis of the best and worst contenders confirms a hypothesis that the more specialized the signal, the better it is for subject identification. Signals that provide common data such as temperature do not offer much for subject identification.

\begin{table}[!htb]
	\centering
	\caption{E-coaching scenario: Subject Identification Performance (\% $\pm$ standard deviation) using Physiological Signals}
	\label{tab:id}
	\begin{footnotesize}
		\begin{tabular}{@{}cc|cc@{}} 
			\multicolumn{2}{c|} {\small Modality} & \small Accuracy & \small $F_1$-score \\
			\hline
			\hline
			\multirow{6}{*}{\small Chest}	&	ACC	&	88.47	$\pm$	2.05	&	88.51	$\pm$	2.02	\\
			&	ECG	&	97.37	$\pm$	1.44	&	97.38	$\pm$	1.41	\\
			&	EDA	&	60.28	$\pm$	2.45	&	57.09	$\pm$	2.77	\\
			&	EMG	&	20.11	$\pm$	4.30	&	13.68	$\pm$	4.81	\\
			&	RESP	&	99.84	$\pm$	0.10	&	99.84	$\pm$	0.11	\\
			&	TEMP	&	19.15	$\pm$	1.68	&	11.04	$\pm$	2.73	\\
			\hline
			\multirow{4}{*}{\small Wrist}	&	ACC	&	96.92	$\pm$	0.71	&	96.92	$\pm$	0.71	\\
			&	BVP	&	99.60	$\pm$	0.19	&	99.60	$\pm$	0.19	\\
			&	EDA	&	51.08	$\pm$	2.75	&	47.65	$\pm$	3.67	\\
			&	TEMP	&	24.03	$\pm$	3.35	&	18.87	$\pm$	3.66	\\
			\hline
		\end{tabular}
	\end{footnotesize}
\end{table}

\subsection{Experiment II: Stressed e-coaching classification}

Once the identity of the wearer is determined, the next step is to perform the general and personalized emotion classification. This is implemented in this paper via leave-one-subject-out cross-validation. This form of cross-validation evaluates the performance of the system when one specific subject's data is never seen by the machine-learning model. The purpose of such evaluation is to analyze the system response to unknown data. The personalized emotion classification is then done via 10-fold cross-validation. 10-fold cross-validation measures the performance of the system when each subject's data is partially shown to the machine-learning model. This validation procedure reports the result when the user knows the identity of the subject.

\begin{table}[!htb]
	\centering
	\caption{E-coaching scenario: Emotion Classification Performance (\% $\pm$ standard deviation) using Physiological Signals: (a) Generalized and (b) Personalized}
	\label{tab:emo}
	\begin{footnotesize}
		\begin{tabular}{@{}c|cc|cc@{}} 
			\multicolumn{1}{c|} {\small}& \multicolumn{2}{c|}{\textbf{\small Generalized Mode}} &
			\multicolumn{2}{c}{\textbf{\small Personalized Mode}}\\ 
			\multicolumn{1}{c|} {Mod.} & \small Accuracy & \small $F_1$-score & \small Accuracy & \small $F_1$-score\\
			\hline
			\hline
			\multicolumn{5}{c}{\small Chest}	\\
			ACC	&	71.2	$\pm$	13.2	&	66.6	$\pm$	15.7	&	84.7	$\pm$	3.8	&	85.2	$\pm$	3.5	\\
			ECG	&	72.7	$\pm$	13.4	&	68.4	$\pm$	16.9	&	92.6	$\pm$	5.7	&	92.6	$\pm$	5.6	\\
			EDA	&	68.6	$\pm$	20.5	&	64.1	$\pm$	24.6	&	60.3	$\pm$	1.8	&	62.0	$\pm$	1.6	\\
			EMG	&	67.9	$\pm$	11.7	&	58.9	$\pm$	14.3	&	56.0	$\pm$	1.4	&	44.9	$\pm$	4.1	\\
			RESP	&	82.9	$\pm$	9.2	&	81.4	$\pm$	9.4	&	99.8	$\pm$	0.1	&	99.8	$\pm$	0.1	\\
			TEMP	&	75.0	$\pm$	10.3	&	67.7	$\pm$	11.6	&	56.8	$\pm$	0.8	&	47.0	$\pm$	2.7	\\
			\hline
			\multicolumn{5}{c}{\small Wrist}	\\
			ACC	&	73.74	$\pm$	16.6	&	72.4	$\pm$	17.4	&	97.3	$\pm$	0.5	&	97.4	$\pm$	0.5	\\
			BVP	&	76.0	$\pm$	10.4	&	72.5	$\pm$	12.6	&	99.6	$\pm$	0.2	&	99.6	$\pm$	0.2	\\
			EDA	&	67.5	$\pm$	18.2	&	63.8	$\pm$	18.9	&	65.5	$\pm$	4.0	&	66.0	$\pm$	4.1	\\
			TEMP	&	59.2	$\pm$	7.1	&	46.3	$\pm$	11.1	&	54.6	$\pm$	1.3	&	42.5	$\pm$	4.6	\\
			\hline
		\end{tabular}
	\end{footnotesize}
\end{table}

Table \ref{tab:emo} reports the emotion classification performance for (a) generalized mode and (b) personalized mode. For each mode, 10 different signals are used for emotion classification, including the accelerometer data, the temperature of the chest, blood volume pulse rate, and electrodermal activity signal measured at the wrist.

\subsubsection*{Observation 2 (Resp- and BVP-centric monitoring)} 
In the general mode, the best-performing signal is respiration (Resp) and BVP for the chest and wrist, respectively. 

\subsubsection*{Observation 3 (Comparison)}
An interesting note is that these results coincide with the identification results. The biggest contrast between Table \ref{tab:id} and Table \ref{tab:emo}(a) is that the TEMP signal performs much better at emotion classification than subject identification. In the personalized mode, it is once again the RESP and BVP signals that offer the highest performance with an accuracy of 99.8\% and 99.6\% for the sensors located at the chest and wrist, respectively.

\subsubsection*{Observation 4 (RESP-centric monitoring)} 
For the RESP signal, the general mode is characterized by an accuracy of 82.9\%. This is boosted to 99.8\% if the identity of the wearer is known.

\subsubsection*{Observation 5 (TEMP-centric monitoring)}
TEMP signal recorded using the chest sensor shows an accuracy of 75.0\% in the general mode, and the accuracy decreases to 56.8\% in the personalized mode.

Signals that provide unique features that can be used for identification can also be used to boost the performance of emotion classification. When there is an absence of identifiable features, the emotion classification performance is detrimentally impacted.

\subsubsection*{Observation 6 (Comparison)}
Comparison between the different signals provides further conclusions. In particular, accelerometer data is a common signal collected by wearable devices and smartphones, while physiological signals such as ECG are not as readily available on smartphones. When we compare the performance between ACC data and other physiological signals, we implicitly compare the performance of the sensor devices. Specifically, the usage of accelerometer data achieves accuracies of 84.7\% and 97.3\% for the chest and wrist sensors, respectively, as shown in Table \ref{tab:emo}.

This performance is comparable to the best-performing signals, Resp and BVP. There is a greater disparity in performance between the Chest-ACC and Chest-Resp, as opposed to the Wrist-ACC and Wrist-BVP. This is most likely the result of accelerometer data being more useful near the hand as opposed to the chest which has a lower degree of movement.

\section{Stage II: Reasoning}
\label{sec:Intelligent-reasoning}

In Section \ref{sec:Proof-of-concept}, deep learning tools were used for stress continuous monitoring. This is the first stage of intelligent processing. The goal of the second stage is to interpret these results using intelligent reasoning tools.   Causal reasoning is a judgment under uncertainty performed on a causal network \cite{[Pearl-2019]}. In this section, we provide two experiments using causal networks.

\subsection{Causal networks }


A recent review \cite{[Rohmer-2020]} describes the various types of causal networks that are deployed in machine reasoning, e.g., Bayesian, imprecise, interval, credal, fuzzy, and subjective networks. The choice of type of causal network depends on the scenario as well as the CUT as a carrier of \emph{primary} knowledge. In our study, among various causal networks, we have chosen to use Bayesian causal networks. Our motivation for this choice is driven by the fact that the Bayesian (probabilistic) interpretation of uncertainty provides acceptable reliability in decision-making. A Bayesian network is defined as a causal network with Conditional Probability Tables (CPTs)  representing point probability measures. 

\begin{figure}[!ht]
	\begin{center}
		\includegraphics[width=0.28\textwidth]{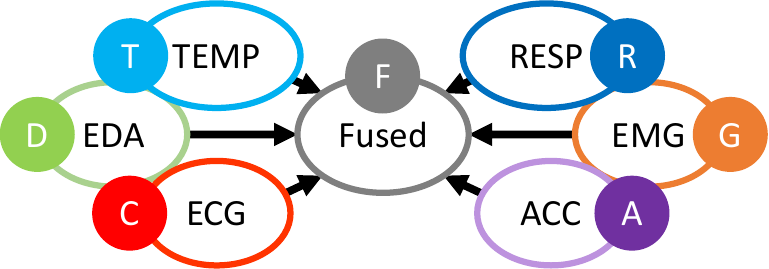}
	\end{center}
	\vspace{-3mm}
	\caption{Causal network with parent nodes being the stress predictors and the child node as a fusion of the predictors.}
	\label{fig:bn}
\end{figure}

Fig. \ref{fig:bn} illustrates a basic causal network containing six parent nodes (representing each of the body sensors) and one child node representing the fusion of separate predictors. 

\subsection{Structural Equation Modeling}
\begin{figure}[!ht]
	\begin{center}
		\includegraphics[clip, trim=1.4cm 1.1cm 1.4cm 1.4cm, width=0.5\textwidth]{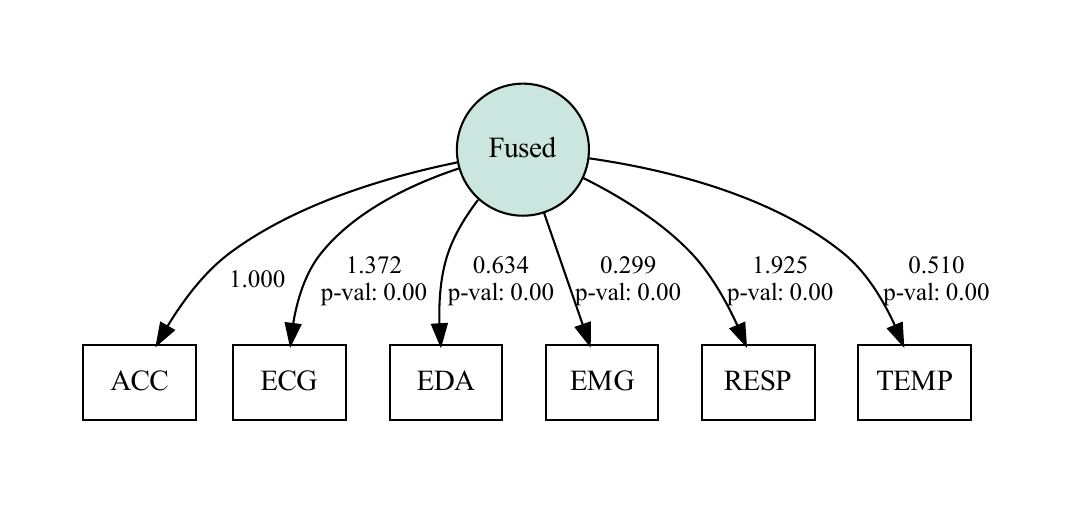}
	\end{center}
	\caption{SEM for personalized prediction of stress using chest sensors.  The values on the arcs represent the regression coefficient and its corresponding $p$-value.  The regression coefficients measure the change between each node and the $p$-values are probabilities used to describe how likely the null hypothesis is true.}
	\label{fig:sem}
\end{figure}

In order to validate the relationship strength between each node in the constructed causal network, we use Structural Equation Modeling (SEM).  In this paper, we used semopy \cite{igolkina2020semopy} to perform structural model analysis.  The objective function used in the SEM is the likelihood function and the optimization method is Sequential Least-Squares Quadratic Programming. 
Fig. \ref{fig:sem} illustrates the estimated parameter and $p$-values between the sensor nodes and the fused node.  The circular node represents the latent variable and the rectangles represent observed/measured variables.  In this paper, we are able to measure the performance of each sensor, reported in terms of accuracy and $F_1$-score.  When every sensor is fused together, the final prediction of stress is dependent on the prediction result of each sensor. Through SEM, we can evaluate which sensor provides the most beneficial prediction of stress.

The regression coefficients in Fig. \ref{fig:sem} indicate a strong relationship to the performance of the classification accuracy in Table \ref{tab:emo}.  We observe that the sensors providing the lowest accuracy such as EMG also have much lower estimated values.  This indicates that specific sensors, such as EDA, EMG, and Temp are not suited to be used, specifically introducing these sensors into the fused results may decrease the combined performance.  The ACC, ECG, and Resp sensors with estimated values of 1, 1.372, and 1.925, respectively provide the best data for personalized prediction of stress.  Thus, the edge weights can be manually altered to focus on better-performing sensors instead of assigning equal weights among all sensors.

\subsection{Experiment III: Stress inference }

\begin{figure}[!ht]
	\begin{center}
		\includegraphics[width=0.45\textwidth]{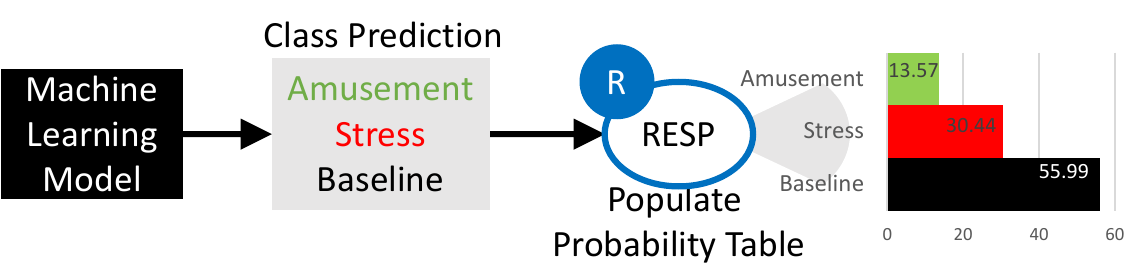}
	\end{center}
	\caption{Emotion class-based CPTs population. CPTs are created using the emotion class predictions provided by the machine-learning model.}\label{fig:bn1_exp}
\end{figure}

In this experiment, we show how a causal network can be used to estimate the expected prediction from a machine-learning model. Accuracy measures the relative performance of the system, e.g., how many predictions rendered by the machine-learning model are correct compared to the ground truth.  
Fig. \ref{fig:bn1_exp} illustrates the process of populating the CPT for node $R$ in the BN shown in Fig. \ref{fig:bn}, using the results of class prediction from a machine-learning model. 

The machine-learning model predicts one of three emotional classes for each test sample, and the prediction statistics for all test cases are collected to generate a discrete distribution for each sensor. This distribution shows what emotion class each sensor will predict, on average, given a random sample. For example, given 100 samples, 59 of the samples are predicted, by the accelerometer node (ACC), as a baseline, 26 as stress, and 15 as amusement. A comparison between each prediction with the ground truth is necessary to know whether the prediction was correct or not. These distributions act as the CPTs for the BN.


  This BN can be used for diagnostics or inference, as shown in the next sections. For example, given an observation that EDA is `baseline', ACC is `stress', and the subject (`Fused') is reported to be `Baseline' (B), through inference, the probabilities for the RESP node are as follows: 61.98\% baseline, 26.24\% stress, and 11.78\% amused.

\subsubsection*{Observation 7 (Amusement monitoring)}

\begin{figure}[!ht]
	\begin{center}
		\includegraphics[width=0.49\textwidth]{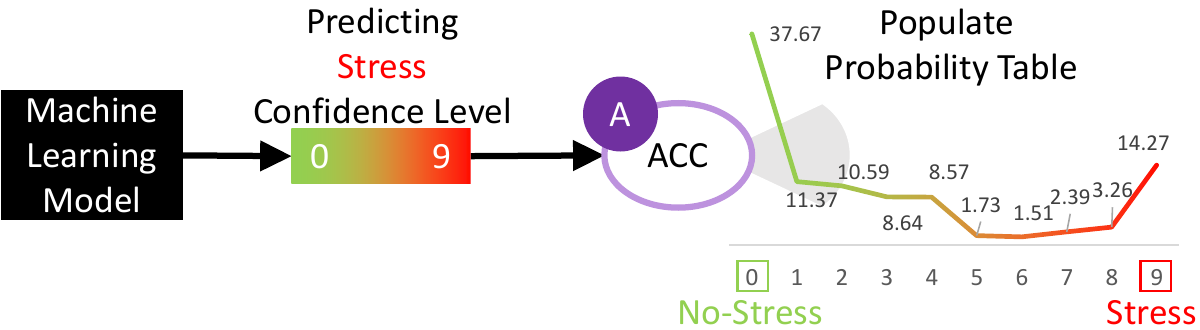}
	\end{center}
	\caption{Stress distribution-based CPT population. CPTs are generated using the stress confidence levels from the output of the machine-learning model.}
	\label{fig:bn2_exp}
\end{figure}


Fig. \ref{fig:bn2_exp} describes the CPTs populating based on the confidence level for the stress level predicted by the machine learning model. The stress confidence level is collected for all test cases and used to create a confidence distribution for each sensor. Each distribution acts as a CPT for the proposed BN.

\subsection{Experiment IV: Stress prediction}

\begin{figure*}[!ht]
	\begin{center}
		\includegraphics[clip, trim=0cm 0.6cm 0cm 0cm, width=0.99\textwidth]{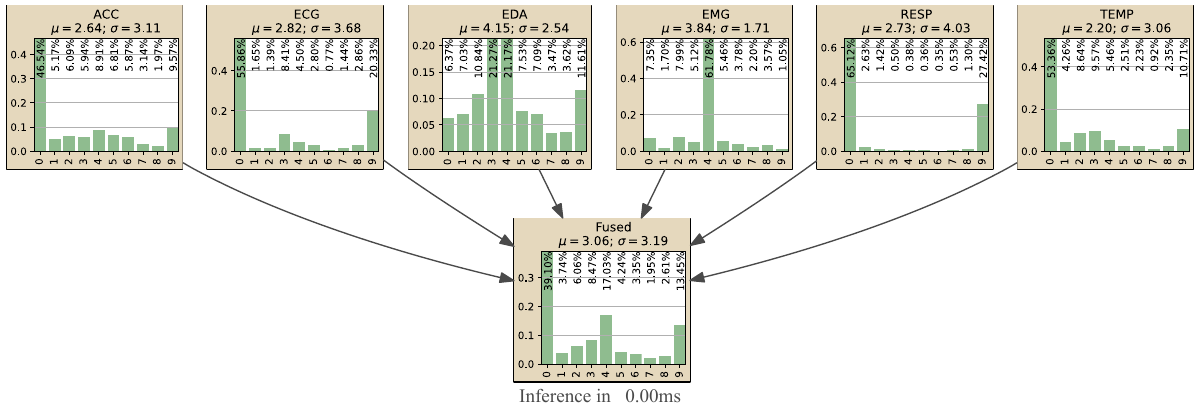}
	\end{center}
	\caption{Stress distribution-based BN example generated using the PyAgrum library. }
	\label{fig:bn2}
\end{figure*}

Fig. \ref{fig:bn2} further adjusts the causal network to account for the distribution of a specific emotion, specifically the confidence or strength in the prediction of the stress emotion. For each node in the network, the $x$-axis represents the confidence of the classifier in prediction where 9 indicates high confidence in detecting stress, 0 indicates high confidence in the absence of stress, and 4-5 indicates that the classifier is not confident in either decision. The distributions in each node report on average what to expect. This is different from accuracy which indicates on average whether a random prediction is correct. For example, consider using the Temperature node (Temp) as a predictor: given 100 samples, 33 samples indicate very high confidence (0) in no-stress, 8 samples render medium confidence (4) in no-stress, and 20 samples suggest high confidence (9) in stress.

\subsubsection*{Observation 8 (ECG and Resp-centric monitoring)}

The relative strength of stress levels for the ECG and Resp sensors tend to be close to either 0 or 9. This observation indicates that both of these sensors are highly confident predictors of no-stress or stress. The overlapping regions between the stress and no-stress decisions are minimal, thus providing two separate distributions. 


\section{Conclusions}\label{sec:Summary}
This study represents the \textbf{first attempt to design the  e-coaching that counts for EMC projections}. Four pillars, i.e., prevention, preparedness, response, and recovery, provide strong systematic requirements for e-coaching during times of emergencies. 
The \textbf{mental state of e-coaching users  became a factor of critical importance}. In previous works on e-coaching, researchers were able only to hypothesize about psychological factors. Nowadays, evidence has accumulated confirming that the stress of the e-coached users is a powerful factor impacting the e-coaching tactic and strategy. 


Our key recommendation to the designers of the e-coaching systems is to include the following \textbf{mandatory mechanisms}:
\begin{enumerate}
	\item \textbf{Taxonomical view}, i.e., e-coaching should be considered as a component of EMC surveillance. 
	\item \textbf{Continuous stress monitoring}, i.e., e-coaching tactic, and, consequently, strategy, must be adjusted once the user's stress state is detected. 
\end{enumerate}

We presented a core idea involving two stages of intelligent processing using deep learning and machine reasoning. Deep learning techniques are used to recognize mental states such as stress. Machine reasoning provides graph-based models that represent the joint distribution of the involved variables such as each sensor's prediction accuracy provided by the machine learning. It also embodies a fusion mechanism, thus, representing each sensor's contribution to the combined system-level decision. These two  tools work hand-in-hand, with machine learning providing each separate variable (sensor) classification accuracy distribution, and machine reasoning for combining those in a joint distribution. Together, these processes can be applied in many different areas including stress monitoring for e-coaching.


%

	\section*{Acknowledgments}
This work was supported in part by the Social Sciences and Humanities Research Council of Canada (SSHRC) through the Grant ``Emergency Management Cycle-Centric R\&D: From National Prototyping to Global Implementation'' under Grant NFRF-2021-00277; in part by the University of Calgary under the Eyes High Postdoctoral Match-Funding Program.

\bibliographystyle{IEEEtran}
\bibliography{test}
	
\end{document}